\documentclass[aps,twocolumn,showpacs,superscriptaddress,groupedaddress,floatfix]{revtex4-1}  
\usepackage{graphicx}  
\usepackage{dcolumn}   
\usepackage{bm}        
\usepackage{amssymb}
\usepackage{amsmath}   
\usepackage{hyperref}

\hypersetup{colorlinks,linkcolor={blue},citecolor={blue},urlcolor={red}}
\newcommand{\Mpl}{M_{_{\rm Pl}}}
\newcommand{\Mp}{M_{_{\rm Pl}}}
\newcommand{\hnl}{h_{_{\rm NL}}}
\newcommand{\pt}{{\mathcal P}_{_{\rm T}}}

\def\d{{\rm d}}
\def\nn{\nonumber} 

\def\pa{{\partial}}
\def\f{\frac}

\def\l{\left}
\def\r{\right}

\def\ei{\eta_{\rm i}}

\def\ee{\eta_{\rm e}}
\def\ef{\eta_{\rm f}}

\def\vk{{\bm k}}

\def\vka{{\bm k}_{1}}
\def\vkb{{\bm k}_{2}}
\def\vkc{{\bm k}_{3}}
\def\ska{{k_{1}}}
\def\skb{{k_{2}}}
\def\skc{{k_{3}}}

\def\cS{\mathcal S}
\def\cL{\mathcal L}
\def\cG{\mathcal G}
\def\cF{\mathcal F}
\newcommand{\g}{\gamma}

\newcommand{\viz}{\textit{viz.~}}
\newcommand{\ie}{\textit{i.e.~}}

\begin{document}
\title{Can non-minimal coupling restore the consistency condition in 
bouncing universes?}
\author{Debottam Nandi}
\email{E-mail: debottam@physics.iitm.ac.in}
\author{L.~Sriramkumar}
\email{E-mail: sriram@physics.iitm.ac.in}
\affiliation{Department of Physics, Indian Institute of Technology Madras, 
Chennai 600036, India}
\begin{abstract}
An important property of the three-point functions generated in the 
early universe is the so-called consistency condition.
According to the condition, in the squeezed limit wherein the wavenumber 
of one of the three modes (constituting the triangular configuration of
wavevectors) is much smaller than the other two, the three-point functions 
can be completely expressed in terms of the two-point functions.
It is found that, while the consistency condition is mostly satisfied by 
the primordial perturbations generated in the inflationary scenario, it 
is often violated in the bouncing models.
The validity of the consistency condition in the context of inflation can 
be attributed to the fact that the amplitude of the scalar and tensor 
perturbations freeze on super-Hubble scales.
Whereas, in the bouncing scenarios, the amplitude of the scalar and tensor 
perturbations often grow rapidly as one approaches the bounce, leading to 
a violation of the condition.
In this work, with the help of a specific example involving the tensor 
perturbations, we explicitly show that suitable non-minimal couplings 
can restore the consistency condition even in the bouncing models.
We briefly discuss the implications of the result.
\end{abstract}
\maketitle


\section{Introduction}

Arguably, the inflationary scenario is the most efficient and 
compelling paradigm to describe the origin of perturbations in 
the early universe~\cite{Mukhanov:1990me,Bassett:2005xm,
Sriramkumar:2009kg,Baumann:2009ds,Sriramkumar:2012mik,
Linde:2014nna,Martin:2015dha}. 
Inflation corresponds to a brief phase of accelerated expansion in 
the early stages of the universe, and it is often invoked to resolve 
the horizon and flatness problems associated with the conventional 
hot big bang model. 
Its success can be attributed primarily to the fact that the simplest 
of inflationary models lead to nearly scale invariant primordial 
spectra which prove to be remarkably consistent with the cosmological 
data~\cite{Akrami:2018odb,Aghanim:2018eyx}. 
However, the impressive efficiency of the inflationary paradigm also 
seems to harbor a possible pitfall. 
Despite the ever-tightening observational constraints, there seems to
exist many inflationary models that continue remain consistent with 
the data~\cite{Martin:2010hh,Martin:2013tda,Martin:2013nzq,Martin:2014rqa},
even leading to the concern whether inflation can be falsified at 
all~\cite{Gubitosi:2015pba}.

\par

A popular alternative to the inflationary paradigm are the classical 
bouncing scenarios~\cite{Novello:2008ra,Cai:2014bea,Battefeld:2014uga,
Lilley:2015ksa,Ijjas:2015hcc,Brandenberger:2016vhg}. 
In these scenarios, the universe undergoes a phase of contraction until 
the scale factor reaches a minimum value, before it enters the expanding 
phase. 
It is straightforward to establish that such scenarios can aid in 
overcoming the horizon problem.
Moreover, it can be shown that a {\it non-accelerating}\/ early phase 
of contraction will permit the standard Bunch-Davies initial conditions 
to be imposed on the perturbations, in the same manner as in inflation. 
However, in complete contrast to inflation, there arises many challenges
in constructing viable bouncing models.
The basic reason behind the difficulty is the fact that the null energy 
condition needs to be violated around the bounce.
Nevertheless, the bouncing models continue to attract constant attention 
in the literature.

\par

In this work, we shall focus on a specific bouncing scenario known 
as the matter bounce~\cite{Starobinsky:1979ty,Finelli:2001sr,
Raveendran:2017vfx}. 
In such scenarios, during the early stages of contraction, the scale 
factor behaves as in a matter dominated universe. 
Also, matter bounces are guaranteed to lead to scale invariant spectra, 
as they are known to be `dual' to de Sitter inflation (in this context,
see Ref.~\cite{Wands:1998yp}).
Though both de Sitter inflation and matter bounces result in similar spectra,
the amplitude and the shape of the three-point functions generated in these
scenarios are expected to be considerably different.
This is because of the difference in the behavior of the evolution of 
the perturbations in these alternative scenarios.
In the context of inflation, it is well known that the amplitude of the 
perturbations freeze on super-Hubble scales.
Due to this reason, the three-point functions generated during inflation
exhibit an interesting property. 
One finds that, in the so-called squeezed limit wherein one of the three
wavenumbers is much smaller than the other two, the inflationary 
three-point functions can be completely expressed in terms of the 
two-point functions, a property that is referred to as the consistency 
condition (see, for example, 
Refs.~\cite{Maldacena:2002vr,Kundu:2013gha,Sreenath:2014nka}).
On the other hand, in most of the bouncing models, the amplitude of the 
perturbations (corresponding to scales of cosmological interest) grow 
rapidly as one approaches the bounce.
Such a behavior results in a violation of the above-mentioned consistency 
condition (in this context, see, for instance, Ref.~\cite{Chowdhury:2015cma}). 

\par

Our goal in this work is to examine whether non-minimal coupling can 
restore the consistency condition in a matter bounce scenario.
Since studying the case of scalars requires considerable modeling, for 
simplicity, we shall focus on the tensor perturbations.
We shall consider a fairly generic scalar-tensor theory~\cite{Horndeski:1974wa,
Deffayet2009,Deffayet:2010qz,Deffayet:2010zh} and work with coupling functions that lead to 
a scale invariant tensor perturbation spectrum in a matter bounce (for 
discussions in the context of inflation, see Refs.~\cite{Gao:2011vs,Gao:2012ib}).
Moreover, we shall work with parameters that result in a tensor amplitude
that is consistent with the current upper bounds on the tensor-to-scalar
ratio $r$ from the observations of the anisotropies in the Cosmic 
Microwave Background (CMB)~\cite{Akrami:2018odb}.
As we shall show explicitly, it is possible to construct non-minimal 
couplings wherein the amplitude of the tensor perturbations freeze soon 
after they leave Hubble radius during the contracting phase and are 
hardly affected by the bounce.
We shall illustrate that the consistency condition is indeed satisfied 
in such cases.

\par

This paper is organized as follows. 
In the following section, we shall briefly describe the non-minimally
coupled model of our interest and arrive at the actions describing the 
tensor perturbations at the second and the third orders.
In Sec.~\ref{sec:ps}, we shall consider a specific form of the non-miminal
coupling and show that the form leads to scale invariant tensor power 
spectrum.
In Sec.~\ref{sec:bs}, we shall evaluate the corresponding tensor bispectrum
and, in Sec.~\ref{sec:hnl}, we shall calculate the tensor non-Gaussianity 
parameter $\hnl$, which is a dimensionless ratio involving the tensor 
bispectrum and power spectrum.
We shall also explicitly show that the consistency condition is satisfied in 
the model.
We shall conclude in Sec.~\ref{sec:d} with a brief discussion. 

\par

A few words on our conventions and notations are in order at this stage
of our discussion.
We shall work with natural units such that $\hbar=c=1$, and we shall 
define the Planck mass to be $\Mpl =(8\,\pi\, G)^{-1/2}$.
We shall adopt the metric signature of $(-, +, +, +)$.
While Greek indices shall denote the spacetime coordinates, Latin
indices shall denote the spatial coordinates, with the exception of $k$ 
which shall be reserved for representing the wavenumber of the 
perturbations.
The overdots and overprimes, as usual, shall denote derivatives with 
respect to the cosmic time~$t$ and the conformal time~$\eta$ associated
with the Friedmann-Lema\^{\i}tre-Robertson-Walker (FLRW) line-element, 
respectively.
Lastly, $a$ shall denote the scale factor and $H$ shall denote the 
Hubble parameter defined as $H={\dot a}/a$.


\section{Model and scenario of interest}\label{sec:m}

We shall consider a theory of gravitation which involves non-minimal
coupling to a scalar field, say, $\phi$.
In four spacetime dimensions, such a theory, in general, can be
described by the action~\cite{Horndeski:1974wa,Deffayet2009,
Deffayet:2010qz,Deffayet:2010zh}
\begin{eqnarray}\label{eq:a}
\cS[g_{\mu\nu},\phi] 
&=& \int \d^4x\, \sqrt{-g}\,\biggl[\cL_2(g_{\mu\nu},X,\phi)
+ \cL_3(g_{\mu\nu},X,\phi)\nn\\
& &+\, \cL_4(g_{\mu\nu},X,\phi) + \cL_5(g_{\mu\nu},X,\phi)\biggr],
\end{eqnarray}
where $X =-(\nabla_{\mu}\phi)^2/2$ denotes the standard kinetic term.
The Lagrangian densities $\mathcal{L}$'s are defined as
\begin{subequations}
\begin{eqnarray}
\cL_2 &=& K(X,\phi),\\\
\cL_3 &=& - G_3(X,\phi)\,\Box\phi,\\
\cL_4 &=& G_4(X,\phi)\,R + G_{4X}(X,\phi)\,\l[(\Box\phi)^2 - 
(\nabla_{\mu\nu}\phi)^2\r],\quad\;\;\;\\
\cL_5 &=& G_5(X,\phi)\, G_{\mu \nu}\, \nabla^{\mu \nu}\phi
- \frac{1}{6} \,G_{5X}(X,\phi)\nn\\
& &\times\,\l[(\Box\phi)^3
-\,3\, \Box\phi\, (\nabla_{\mu \nu}\phi)^2
+ 2\, (\nabla_{\mu \nu}\phi)^3\r],
\end{eqnarray}
\end{subequations}
where the quantities $K(X,\phi)$ and $G(X,\phi)$ are general functions 
of~$X$ and $\phi$, while the subscript $X$ denotes derivative of the 
function with respect to~$X$. 
Note that the action~(\ref{eq:a}) contains second time derivatives of 
the field~$\phi$. 
Hence, one may naively expect that the corresponding equations of motion
may involve higher time derivatives.
However, the complete action has been structured in a fashion such that 
the governing equations do not contain any higher time derivatives than 
the second.
Therefore, the model is free of the so-called Ostrogradsky 
instabilities~\cite{Ostro}.

\par 

As we had mentioned, in this work, we shall be focusing on the three-point
function describing the tensor perturbations.
When the tensor perturbations, say, $\g_{ij}(\eta,{\bm x})$, are taken into 
account, the spatially flat, FLRW line-element can be written as
\begin{equation}
\d s^2 = a(\eta)^2\,\l[- \d\eta^2
+ \l\{{\rm exp}\,{\g(\eta,{\bm x}})\r\}_{ij}\, 
\text{d}x^i\, \text{d}x^j\r],\label{eq:pfu}
\end{equation}
where, evidently, $a(\eta)$ denotes the scale factor, with~$\eta$ being
the conformal time coordinate.
The quantity $\l\{{\rm exp}\,{\g(\eta,{\bm x}})\r\}_{ij}$ contains the 
tensor perturbations and is defined as 
\begin{equation}
\l\{{\rm e}^\g\r\}_{ij}
= \delta_{ij} + \g_{ij} 
+ \f{1}{2}\,\g_i^l\, \g_{lj} 
+ \f{1}{6}\,\g_i^l\, \g_l^m\, \g_{mj} + \cdots,
\end{equation} 
where the spatial indices are to be raised and lowered with the aid of
Kronecker~$\delta_{ij}$.

\par

Since we shall be focusing on the tensor perturbations, we shall be
interested in only the following part of the action~(\ref{eq:a}):
\begin{eqnarray}
\cS[g_{\mu\nu}] = \int \d^4x\, \sqrt{-g}\,
\l[\cL_4(g_{\mu\nu},X,\phi) + \cL_5(g_{\mu\nu},X,\phi)\r].\nn\\
\label{eq:aoi}
\end{eqnarray}
Upon substituting the line-element~(\ref{eq:pfu}) in this action, 
one can arrive at the following actions that describe the tensor 
perturbations~$\g_{ij}$ at the second and the third order in the 
perturbations (in this context, see Refs.~\cite{Gao:2011vs,Gao:2012ib}):
\begin{subequations}\label{eq:gstoa}
\begin{eqnarray}
\delta^2\cS[\g_{ij}] 
&=& \f{1}{8}\,\int\d\eta\,\int\d^3{\bm x}\,
a^2(\eta)\,
\biggl[\cG(\eta)\,\g_{ij}^{\prime}{}^2\nn\\
& &-\, \cF(\eta)\,\l(\pa_l \g_{i j}\r)^2\biggr],\label{eq:gsoa}\\
\delta^3\cS[\g_{ij}] 
&=& \f{1}{4}\,\int\d\eta\int\d^3{\bm x}\,
\biggl[a^2(\eta)\,\cF(\eta)\,\pa_{lm}\g_{ij}\nn\\
& &\times\,\l(\g^{lj}\,\g^{im}
- \f{1}{2}\,\g^{ij}\, \g^{lm}\r)\nn\\
& &+\, \frac{1}{3}\,X\,\phi^\prime\,G_{5X}(\eta)\, 
\g^i_j{}'\,\g^j_l{}'\,\g^l_i{}'\biggr].\label{eq:gtoa}
\end{eqnarray}
\end{subequations}
The functions $\cG$ and $\cF$ are background quantities and 
are given by
\begin{subequations}\label{eq:cFcG}
\begin{eqnarray}
\cF &=& 2\, \l[G_4 - X\,\l(\ddot{\phi}\,G_{5X} 
+ G_{5\phi}\r)\r],\\
\cG &=& 2\, \l[G_4 - 2\,X\,G_{4X} - X\,
\l(H\,\dot{\phi}\,G_{5X} - G_{5\phi}\r)\r],\quad\;\;
\end{eqnarray}
\end{subequations}
where the subscript $\phi$ denotes differentiation with respect 
to the scalar field.

\par

Note that the standard Einstein's general theory of relativity
corresponds to choosing
\begin{subequations}
\begin{eqnarray}
G_4(X,\phi) &=& \f{\Mpl^2}{2},\\
G_5(X,\phi) &=& 0.
\end{eqnarray}
\end{subequations}
In such a case, the general second and third order actions~(\ref{eq:gstoa})
that describe the tensor perturbations reduce to the following forms
(see, for instance, Refs.~\cite{Maldacena:2002vr,Sreenath:2014nka,
Chowdhury:2015cma}):
\begin{subequations}
\begin{eqnarray}
\delta^2\cS[\g_{ij}] 
&=& \f{\Mpl^2}{8}\,\int\d\eta\,\int\d^3{\bm x}\,a^2(\eta)\,
\l[\g_{ij}^{\prime}{}^2 - \l(\pa_l \g_{i j}\r)^2\r],\nn\\
\\
\delta^3\cS[\g_{ij}]  
&=& \f{\Mpl^2}{4}\,\int\d\eta\,\int\d^3{\bm x}\,a^2(\eta)\,
\partial_{lm}\g_{ij}\nn\\
& &\times\,\l(\g^{lj}\,\g^{im} - \f{1}{2}\,\g^{i j}\, \g^{lm}\r).
\end{eqnarray}
\end{subequations}
In an earlier work, assuming the matter bounce to be described by the 
scale factor 
\begin{equation}
a(\eta) =  a_0\,(1 + k_0^2\, \eta^2)
= a_0\,\l(1 + \f{\eta^2}{\eta_0^2}\r),\label{eq:mb} 
\end{equation}
the tensor power and bispectra were evaluated analytically in Einstein's 
theory using the above second and third order actions (in this context,
see Ref.~\cite{Chowdhury:2015cma}).
Clearly, in the above scale factor, $\eta_0=1/k_0$, and the duration of
the bounce (in terms of cosmic time) is of the order of $a_0\,\eta_0$.
Moreover, the wavenumbers $k$ of cosmological interest correspond to $k
\ll k_0$.
The tensor modes for such wavenumbers were obtained under a certain 
approximation and they were evolved across the bounce (at $\eta=0$) 
to arrive at the power and bispectra after the bounce.
Since the matter bounce is dual to de Sitter inflation, as expected, 
the tensor power spectrum proved to be strictly scale invariant for 
modes of cosmological interest. 
Interestingly, the power spectrum had depended only on the dimensionless
ratio $k_0/(a_0\,\Mpl)$ and, for $k_0/(a_0\,\Mpl)\lesssim 10^{-5}$, the 
tensor amplitude had proved to be consistent with the current constraints 
from the CMB data~\cite{Akrami:2018odb}.
Having arrived at the modes, it was also possible to evaluate the tensor
bispectrum and the corresponding non-Gaussianity parameter~$\hnl$.
The amplitude and shape of the tensor bispectrum and the non-Gaussianity 
parameter had turned out to be considerably different from what arises in 
de Sitter inflation.
For instance, while the tensor non-Gaussianity parameter $\hnl$ is
found to be strictly scale invariant in the equilateral and the 
squeezed limits in de Sitter inflation, the parameter had a strong 
dependence on the wavenumber (it had behaved as $k^2$) in the matter 
bounce scenario.
The strong dependence of the non-Gaussianity parameter on the wavenumber
also led to considerably smaller values for the parameter (when compared
to the de Sitter case) over scales of cosmological interest.
Moreover, it was found that the consistency condition is violated 
in the squeezed limit~\cite{Chowdhury:2015cma}.
As we have already emphasized, such differences in the three-point 
functions are expected to help us discriminate between the alternative
scenarios for the generation of perturbations in the early universe.

\par

The differences between the tensor bispectra in de Sitter inflation and
the matter bounce scenarios arise due to the behavior of the modes in 
these two cases~\cite{Maldacena:2002vr,Chowdhury:2015cma}.
While in de Sitter, as we have already discussed, the amplitude of the 
tensor modes freeze once they leave the Hubble radius, in the matter 
bounce scenario, the amplitude of the modes grow rapidly as they approach 
the bounce after leaving the Hubble radius during the contracting phase.
It is then interesting to inquire if, in non-minimally coupled theories
of gravitation, the tensor perturbations can behave in a manner akin to
de Sitter and thereby restore the consistency condition in the bouncing 
models. 

\par

With such a motivation in mind, we make the following choices for the 
functions $G_4(X,\phi)$ and  $G_5(X,\phi)$ that appear in the 
non-minimal action~(\ref{eq:aoi}):
\begin{subequations}
\begin{eqnarray}
G_4(X,\phi) &=& \f{\Mpl^2}{2}\,G(\phi),\\
G_5(X,\phi) &=& {\rm constant}.
\end{eqnarray}
\end{subequations}
In such a case, since the derivatives $G_{4X}$, $G_{5X}$ and $G_{5\phi}$ 
vanish, we find that [cf. Eqs.~(\ref{eq:cFcG})]
\begin{equation}
\cG=\cF=2\, G_4(\phi)=\Mpl^2\,G(\phi).
\end{equation}
Then, the second and third order actions~(\ref{eq:gstoa}) describing the 
tensor perturbations reduce to 
\begin{subequations}
\begin{eqnarray}
\delta^2\cS[\g_{ij}] 
&=& \frac{\Mpl^2}{8}\,\int\d\eta\,\int\d^3{\bm x}\,z^2(\eta)\,
\l[\g_{ij}^{\prime}{}^2 - \l(\partial_l \g_{i j}\r)^2\r],\label{eq:soaoi}\nn\\
\\
\delta^3\cS[\g_{ij}] 
&=& \frac{\Mpl^2}{4}\,\int\d\eta\,\int\d^3{\bm x}\,z^2(\eta)\,
\partial_{lm}\g_{ij}\nn\\
& &\times\,\l(\g^{lj}\,\g^{im} - \f{1}{2}\,\g^{i j}\, \g^{lm}\r),
\label{eq:toaoi}
\end{eqnarray}
\end{subequations}
where we have set 
\begin{equation}
z^2=a^2\,G.
\end{equation}
Evidently, the structure of the above set of actions have the same 
form as in Einstein's theory with the overall factor $a(\eta)$ being 
replaced by the function $z(\eta)$, which is determined by~$G(\phi)$. 
We shall choose to work with 
\begin{equation}
G(\eta)=\f{C^2}{a^3(\eta)},\label{eq:nmc}
\end{equation}
where $C$ is a dimensionless constant. 
Such a choice leads to 
\begin{equation}
z(\eta)=\f{C}{\sqrt{a(\eta)}},\label{eq:z}
\end{equation}
and we shall discuss the reason for this choice in the final section.
In the following two sections, we shall evaluate the tensor power and 
bispectra spectra for the above choice of~$z(\eta)$ assuming that 
$a(\eta)$ describes a matter bounce as in Eq.~(\ref{eq:mb}).


\section{Evolution of the tensor modes and the power spectrum}\label{sec:ps}

In this section, we shall arrive at the solutions to the Fourier modes
describing the tensor perturbations and evaluate the tensor 
perturbation spectrum {\it after}\/ the bounce.

\par

\begin{figure}[!t]
\begin{center}
\includegraphics[width=8.60cm]{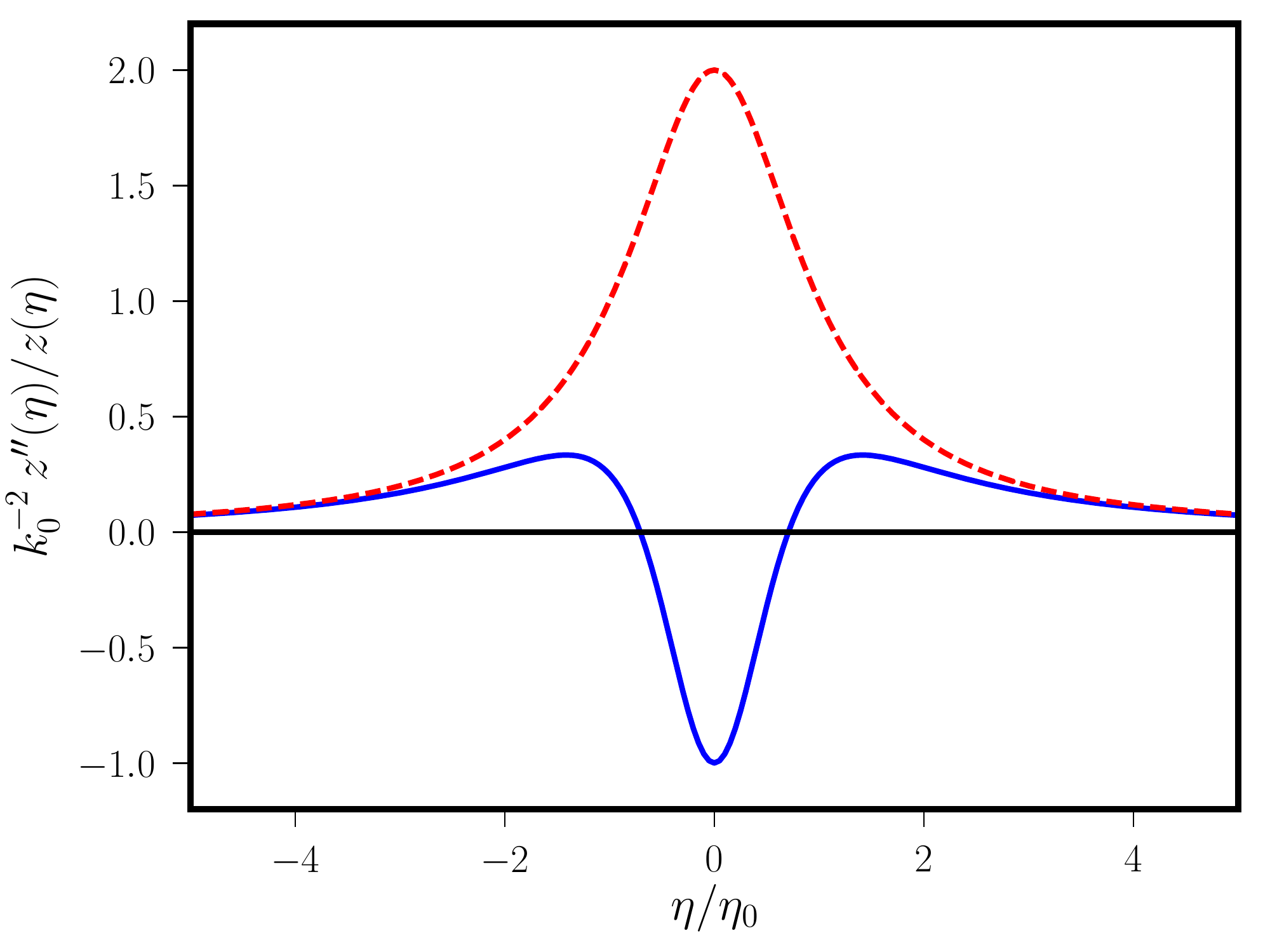} 
\end{center}
\caption{The dimensionless quantity $k_0^{-2}\,(z''/z)$ has been plotted 
(in blue) as a function of $\eta/\eta_0$.
We should point out that none of the parameters involved (\viz $C$, $k_0$ 
or $a_0$) need to be specified in plotting this figure.
Note that the quantity $k_0^{-2}\,(z''/z)$ grows as one approaches the 
bounce, but exhibits a minimum at the bounce.
The minimum is absent when $z=a$ (plotted in red), which is the case in 
Einsteinian gravity~\cite{Chowdhury:2015cma}.}\label{fig:zppz}
\end{figure}

Let $h_k$ denote the Fourier modes corresponding to the tensor 
perturbations $\g_{ij}$.
Upon varying the action~(\ref{eq:soaoi}) with respect to $\g_{ij}$, 
one finds that the corresponding Fourier modes $h_k$ satisfy the 
differential equation
\begin{equation}
h_k''+2 \,\f{z''}{z}\,\,h_k' + k^2\,h_k = 0.
\end{equation}
If we introduce the Mukhanov-Sasaki variable $u_k=(\Mpl/\sqrt{2})\,
h_k\,z$, this equation takes the form 
\begin{equation}
u_k''+ \l(k^2 - \f{z''}{z}\right)\,u_k = 0.
\end{equation}
For $z(\eta)=C/\sqrt{a(\eta)}$ and the scale factor~(\ref{eq:mb}), 
one finds that 
\begin{equation}
\f{1}{k_0^2}\,\frac{z''}{z}
=\f{-1 + 2\, k_0^2\, \eta^2}{(1 + k_0^2\, \eta^2)^2}.
\end{equation}
We have plotted the behavior of $z''/z$ in Fig.~\ref{fig:zppz}.
It turns out to be difficult to obtain an exact analytical solution 
for~$u_k$ (or, equivalently, $h_k$) for such a $z''/z$.
However, one can obtain remarkably accurate solutions for the tensor modes 
of cosmological interest under certain approximations.

\par

Note that $z''/z\to 0 $ as $\eta\to -\infty$.
Therefore, the standard Bunch-Davies initial conditions can be imposed 
on the modes at such early times.
Our aim is to evolve the modes from such initial conditions across the 
bounce and evaluate the tensor perturbations after the bounce at, say, 
$\eta=\beta\,\eta_0$, where $\beta$ is suitably large positive number.
In order to arrive at analytical solutions for the modes, we shall divide 
the period of our interest, \viz $-\infty<\eta<\beta\,\eta_0$, into two 
domains, say, $-\infty<\eta<-\alpha\,\eta_0$ and $-\alpha\,\eta_0<\eta<
\beta\,\eta_0$, where we shall again choose $\alpha$ to be a large 
positive number.
We shall indicate possible values for $\alpha$ and $\beta$ in due course.

\par

\begin{figure}[!b]
\begin{center}
\includegraphics[width=8.60cm]{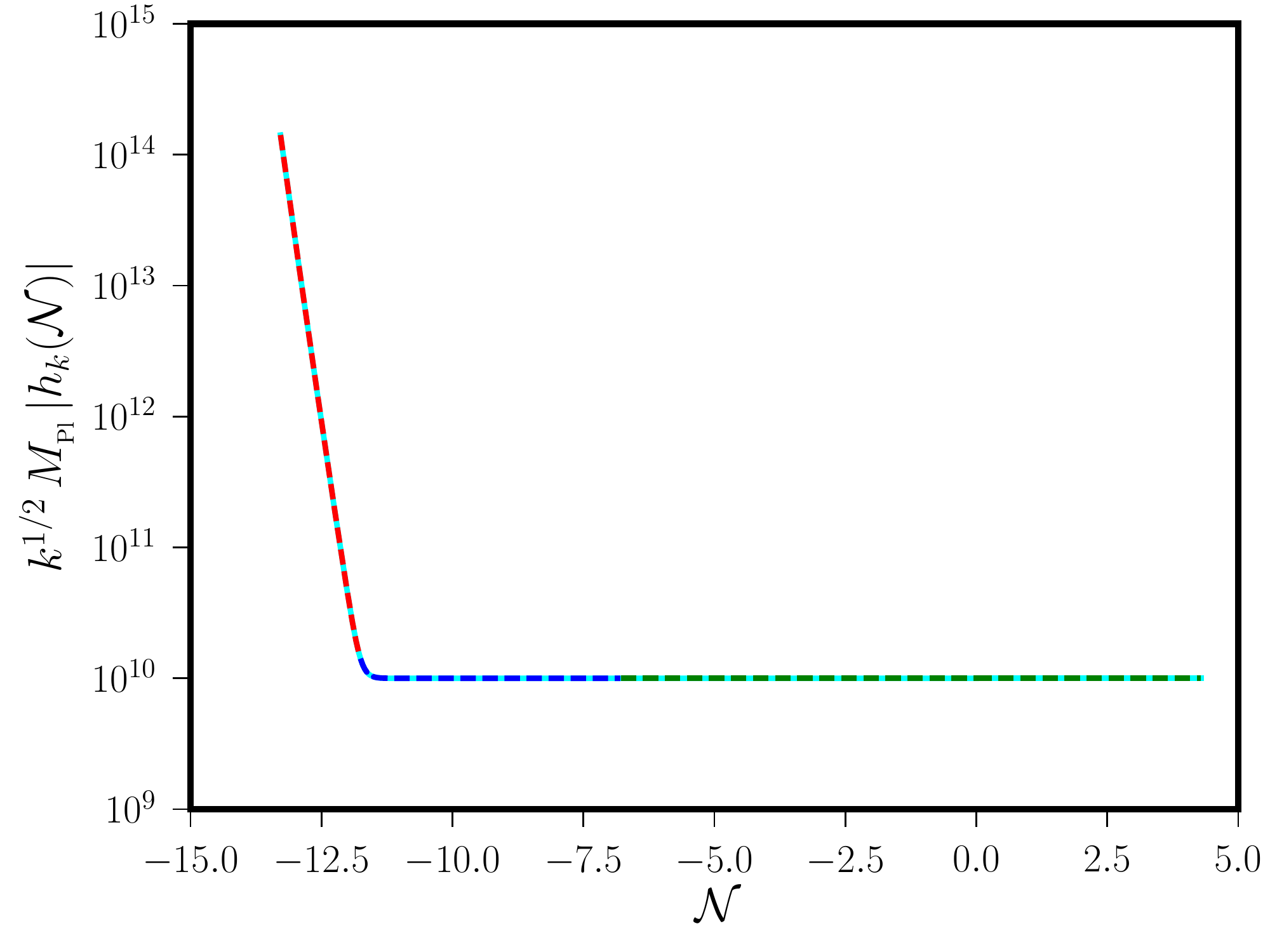} 
\end{center}
\caption{The evolution of the amplitude of the tensor mode~$h_k$ 
evaluated analytically (in red until $\eta_{\rm e}\simeq -\sqrt{2}/k$, 
in blue over $\eta_{\rm e}<\eta<\eta_\ast=-\alpha\,\eta_0$ and in green 
over $\eta_\ast <\eta<\eta_{\rm f}=\beta\,\eta_0$) and numerically (in 
cyan) has been plotted as a function of e-N-folds ${\cal N}$, which is 
defined as $a({\cal N})=a_0\,\exp\,({\cal N}^2/2)$ (in this context,
see Refs.~\cite{Sriramkumar:2015yza,Chowdhury:2015cma}.
Note that the bounce occurs at ${\cal N}=0$, with negative and positive
values of ${\cal N}$ corresponding to the contracting and expanding phases,
respectively.
We have set $k_0/(a_0\,\Mpl)= 10$, $\alpha=10^5$ and $\beta=10^2$ in 
plotting the figure.
The mode of interest corresponds to the wavenumber $k=10^{-11}\,(k_0/\alpha)$, 
which satisfies the required condition $k/(k_0/\alpha)\ll 1$ for the
analytical approximations to be valid.
It is clear that the analytical results match the exact numerical results
very well.
Also, evidently, the amplitude of the tensor modes freeze soon after they 
cross the Hubble radius (to be precise, when $k^2\simeq z''/z$ at~$\ee$) 
during the contracting phase, and remain largely unaffected by the 
bounce.}\label{fig:eh}
\end{figure}

In the first domain, \ie over $-\infty<\eta<-\alpha\,\eta_0$, the scale
factor can be approximated as $a(\eta)\simeq a_0\,k_0^2\,\eta^2$.
In such a case, we have $z(\eta)\simeq -C/(\sqrt{a_0}\,k_0\,\eta)$ and, 
hence, $z''/z\simeq 2/\eta^2$, which is exactly the behavior in de Sitter.
Since $u_k$ corresponding to the Bunch-Davies initial conditions is known 
in this case, the solution to the tensor modes~$h_k$ can be immediately
written down to be~\cite{1978RSPSA.360..117B}
\begin{eqnarray}
h_k^{\rm I}(\eta)
&=&\f{\sqrt{2}}{\Mpl}\, \frac{u_k(\eta)}{z(\eta)}\nn\\ 
&\simeq& -\f{\sqrt{2}}{\Mpl}\,
\f{\sqrt{a_0}\,k_0\,\eta}{C}\,\f{1}{\sqrt{2\,k}}\,
\l(1- \f{i}{k\,\eta}\r)\,{\rm e}^{-i\,k\,\eta}.\label{eq:hkI}\quad
\end{eqnarray}
As we had mentioned earlier, scales of cosmological interest correspond
to $k\ll k_0$.
Note that, during the contracting phase, for such wavenumbers, the solution
$h_k^{\rm I}$ above oscillates over the domain wherein $k^2>z''/z\simeq 
2/\eta^2$. 
Needless to add, it is this oscillating behavior that permits one to 
impose the required initial conditions on the modes.
The amplitude of the modes begin to freeze around the time when $k^2 = z''/z 
\simeq 2/\eta^2$, \ie at  $\eta_{\rm e}\simeq -\sqrt{2}/k$, which closely 
corresponds to the time when the modes exit the Hubble radius.
Thereafter, the amplitude of the modes remain constant until one 
approaches close to the bounce (in this context, see Fig.~\ref{fig:eh}).

\par

Note that the quantity $z''/z$, though it grows to a maximum value 
[of ${\mathcal O}(k_0^2)$] as one approaches the bounce, it also 
contains a minimum [of ${\mathcal O}(-k_0^2)$] at the bounce.  
In the minimally coupled model wherein the tensor modes are governed
only by the behavior of the scale factor, such a minimum is absent, and 
hence it is justified to assume that $k^2\ll a''/a$ around the bounce.
However, in the situation of our interest, we find that $k^2 = z''/z$
at $\eta_{\rm c}=\mp \eta_0/\sqrt{2}$.
Close to these points, evidently, the conditions $k^2\ll z''/z$ will 
not be satisfied.
We shall nevertheless assume this condition is indeed satisfied and 
evaluate the modes across the bounce. 
We shall compare our analytical solutions with the numerical results
to justify our assumption. 
We shall find that, since the period over which the condition is violated 
is rather brief (only very near $\eta=\eta_{\rm c}$), the corresponding 
effects on the modes prove to be completely negligible.
If we now assume that $k^2\ll z''/z$ around the bounce, we can obtain
the solution in the second domain (\ie over $-\alpha\,\eta_0<\eta<
\beta\,\eta_0$) to be  
\begin{eqnarray}
h_k^{\rm II}(\eta) 
= A_k  + B_k\,\l(k_0\,\eta 
+ \frac{k_0^3\,\eta^3}{3}\r),\label{eq:hkII}
\end{eqnarray}
where the constants $A_k$ and $B_k$ can be expressed in terms of 
$h_k(\eta)$ in the first domain as follows:
\begin{subequations}\label{eq:AkBk} 
\begin{eqnarray}
B_k &=& \f{h_k^{\rm I}{}'(\eta_\ast)}{k_0\,\l[a(\eta_\ast)/a_0\r]},\\
A_k &=& h_k^{\rm I}(\eta_\ast) 
- B_k\,\l(k_0\,\eta_\ast + \frac{k_0^3\,\eta_\ast^3}{3}\r).
\end{eqnarray}
\end{subequations}
We can now choose $\eta_\ast= -\alpha\,\eta_0$ to be the time at 
which we match the solution and its time derivative in the two
domains.
It is then easy to determine $A_k$ and $B_k$ to be
\begin{subequations}
\begin{eqnarray}
B_k&=&-\f{\sqrt{2}}{\Mpl}\,\f{\sqrt{a_0}}{C}\,
\f{1}{\sqrt{2\,k}}\,\l(\f{i\,k}{\alpha\,k_0}\r)\,
{\rm e}^{i\,\alpha\,k/k_0},\\
A_k&=&\f{\sqrt{2}}{\Mpl}\,\f{\sqrt{a_0}\,\alpha}{C}\,
\f{1}{\sqrt{2\,k}}\,\l(1+\f{i\,k_0}{\alpha\,k}\r)\,
{\rm e}^{i\,\alpha\,k/k_0}\nn\\
& &+\,B_k\,\l(\alpha+\f{\alpha^3}{3}\r).
\end{eqnarray}
\end{subequations}
For cosmological scales, we expect that $\ee \simeq
-\sqrt{2}/k \ll -\alpha\,\eta_0 =-\alpha/k_0$, which 
translates to the condition $k\ll k_0/\alpha$.
It is for such wavenumbers that our approximation works well.
In Fig.~\ref{fig:eh}, we have plotted the analytical and the 
numerical solution for the tensor modes corresponding to a 
wavenumber that satisfies the above condition.
It should be obvious from the figure that the analytical solution matches 
the numerical solution quite well, indicating the extent of accuracy of 
the analytical approximation.
Also, note that the amplitude of the tensor mode freezes soon after
it exits the Hubble radius during the contracting phase. 
Moreover, it is clear that the bounce does not affect its amplitude.
In other words, the tensor modes broadly behave in a fashion similar 
to the way they do in de Sitter inflation.

\par

\begin{figure}[!t]
\begin{center}
\includegraphics[width=8.60cm]{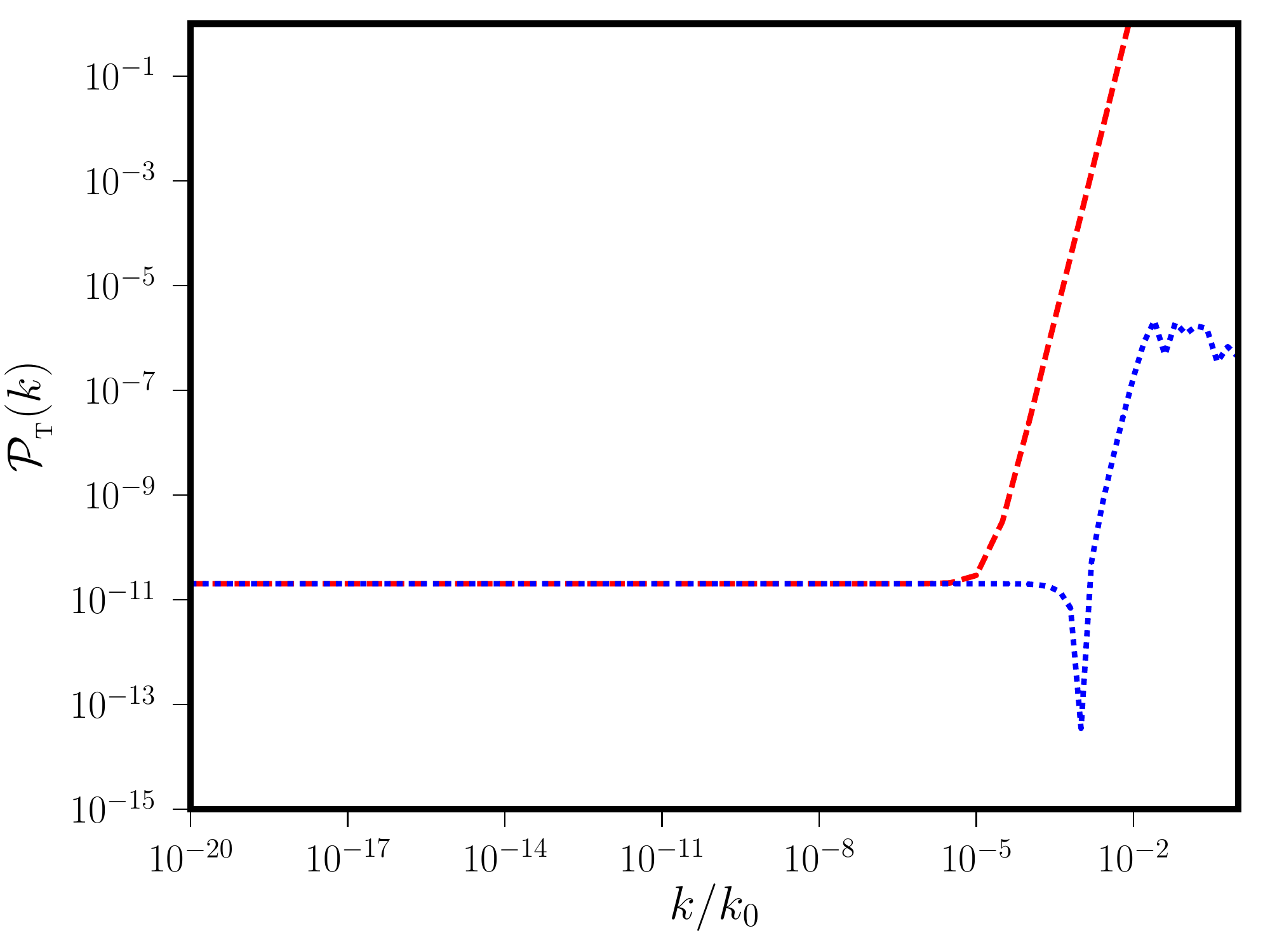} 
\end{center}
\vskip -15pt
\caption{The tensor power spectrum evaluated analytically 
[using Eq.~(\ref{eq:atps})] (in red) and numerically (in blue) 
has been plotted as a function of $k/k_0$.
We have worked with the same set of values of $k_0/(a_0\,\Mpl)$, 
$\alpha$ and $\beta$ as in the previous figure.
Note that the analytical approximations are expected to be 
valid for $k/k_0 \ll 1/\alpha$, a domain over which the 
tensor power spectrum is strictly scale invariant.
These values correspond to the analytical estimate of [cf.
Eq.~(\ref{eq:ata})] $\pt(k)\simeq 2.026\times10^{-11}$ 
over the scale invariant domain, which is exactly what we 
obtain numerically.
Needless to add, the analytical estimates are in very good agreement 
with the numerical results, indicating the validity of the approximations 
involved.}\label{fig:pt}
\end{figure}

Let us now turn to the evaluation of the power spectrum after the 
bounce.
Recall that the tensor power spectrum $\pt(k)$, evaluated at a 
given time, is defined as
\begin{equation}
\pt(k) = 4\,\f{k^3}{2\,\pi^2}\,\vert h_k(\eta)\vert^2.
\end{equation}
Upon using the solution~(\ref{eq:hkII}) in the second domain, the 
tensor power spectrum, evaluated at $\eta_{\rm f}=\beta\,\eta_0$
can be expressed as 
\begin{eqnarray}
\pt(k)=4\,\f{k^3}{2\,\pi^2}\, 
\l\vert A_k+B_k\,\l(\beta+\f{\beta^3}{3}\r)\r\vert^2.\label{eq:atps}
\end{eqnarray}
Clearly, it would desirable that the non-miminal action~(\ref{eq:aoi}) 
reduces to the Einstein-Hilbert form at $\ef$.
This implies that we can set $C = a^{3/2}(\ef)$.
Also, we expect $z$ to behave as $z=a$ thereafter.
In such a case, we find that, for $k\ll k_0/\alpha$, the power spectrum 
proves to be strictly scale invariant and has the following amplitude 
(if we assume that~$\alpha\gg \beta$): 
\begin{eqnarray}
\pt(k)=\f{2}{\pi^2\,\beta^6}\,\l(\f{k_0}{a_0\,\Mpl}\r)^2.\label{eq:ata}
\end{eqnarray}
If we now choose that $k_0/(a_0\,\Mpl)< (\pi\,\beta^3/\sqrt{2})\,
10^{-5}$, one finds that $\pt(k)<10^{-10}$, which will be 
consistent with the current upper bound on the tensor-to-scalar 
ratio of $r\lesssim 0.07$ from the CMB~\cite{Akrami:2018odb}.
In Fig.~\ref{fig:pt}, we have plotted the analytical result~(\ref{eq:atps})
for the tensor power spectrum as well as the corresponding numerical
results.
Evidently, the analytical and numerical results match well for 
$k/k_0\ll 1/\alpha$.

\begin{widetext}


\section{The tensor bi-spectrum}\label{sec:bs}

Given the third order action~(\ref{eq:toaoi}), the corresponding 
interaction Hamiltonian can be easily arrived at 
(see Ref.~\cite{Chen:2006nt}; also see 
Refs.~\cite{Nandi:2015ogk,Nandi:2016pfr}).
With the interaction Hamiltonian at hand, the tensor bispectrum 
can be obtained by using the conventional rules of perturbative
quantum field theory.
The tensor bispectrum $G_{\g\g\g}^{m_1n_1m_2n_2m_3n_3}(\vka,\vkb,\vkc)$, 
calculated in the perturbative vacuum, can be expressed in terms of the 
modes $h_k$ as follows~\cite{Maldacena:2002vr,Maldacena:2011nz,Gao:2011vs,
Gao:2012ib,Sreenath:2013xra,Sreenath:2014nka}:
\begin{eqnarray} \label{eq:Gggg}
G_{\g\g\g}^{m_1n_1m_2n_2m_3n_3}(\vka,\vkb,\vkc)
&= & \Mp^2\; \biggl[\biggl(\Pi_{m_1n_1,ij}^{\vka}\,\Pi_{m_2n_2,im}^{\vkb}\,
\Pi_{m_3n_3,lj}^{\vkc}
-\f{1}{2}\;\Pi_{m_1n_1,ij}^{\vka}\,\Pi_{m_2n_2,ml}^{\vkb}\,
\Pi_{m_3n_3,ij}^{\vkc}\biggr)\, k_{1m}\, k_{1l}\nn\\
& &+\, {\rm five~permutations}\biggr]\,
\biggl[h_{\ska}(\ee)\, h_{\skb}(\ee)\, h_{\skc}(\ee)\,
\cG_{\g\g\g}(\vka,\vkb,\vkc)
+ {\rm complex~conjugate}\biggr],\nn\\\label{eq:tbs-ov}
\end{eqnarray}
\end{widetext}
where $\Pi_{ij,mn}^{\vk}=\sum_{s=1,2} \varepsilon_{ij}^{s}(\vk)\,
\varepsilon_{mn}^{s\ast}(\vk)$, with $\varepsilon_{ij}^{s}(\vk)$
being the polarization tensor characterizing the perturbations
corresponding to helicity~$s$.
The quantity $\cG_{\g\g\g}(\vka,\vkb,\vkc)$ is described by the 
integral
\begin{equation}
\cG_{\g\g\g}(\vka,\vkb,\vkc)
=-\f{i}{4}\,\int_{\ei}^{\ef} \d\eta\; z^2\, h_{\ska}^{\ast}\,
h_{\skb}^{\ast}\,h_{\skc}^{\ast},\label{eq:cG}
\end{equation}
with $\ei$ being the time when the initial conditions are imposed on the 
perturbations and $\ef$ denoting the final time when the bispectrum is to 
be evaluated.
Also, we should clarify that $(k_{1i},k_{2i}, k_{3i})$ denote the 
components of the three wavevectors $({\bm k}_1, {\bm k}_{2}, 
{\bm k}_{3})$ along the $i$-spatial direction.

\par

Our primary aim in this work is to calculate the magnitude and shape of 
the tensor bispectrum and the corresponding non-Gaussianity parameter in
the model of our interest and compare them with, say, the results 
in de Sitter inflation.
Therefore, for convenience, we shall set the polarization 
tensor~$\varepsilon_{ij}^{s}(\vk)$ to unity.
In such a case, the expression~(\ref{eq:tbs-ov}) for the tensor bi-spectrum 
above reduces to the following simpler form:
\begin{eqnarray}
G_{\g\g\g}(\vka,\vkb,\vkc)
&=& \Mp^2\,\biggl[h_{\ska}(\ef)\, h_{\skb}(\ef)\,h_{\skc}(\ef)\nn\\
& &\times\,\l(k_1^2+k_2^2+k_3^2\r)\,\cG_{\g\g\g}(\vka,\vkb,\vkc)\nn\\
& &+\, {\rm complex~conjugate}\biggr].\label{eq:tbs}
\end{eqnarray}
We shall choose $\ei=-\infty$, \ie the earliest time during the contracting 
phase when the initial conditions are imposed on the modes, and we shall 
set~$\ef=\beta\,\eta_0$, \viz the time after the bounce at which we had 
evaluated the power spectrum.

\par

\begin{widetext}
Since we had divided the domain of interest into two to arrive at the 
tensor modes, we can evaluate the quantity $\cG_{\g\g\g}(\vka,\vkb,\vkc)$
[cf. Eq.~(\ref{eq:cG})] over these two domains as well.
Recall that, in the first domain wherein $-\infty< \eta < - \alpha\, \eta_0$, 
we have $z\simeq -C/(\sqrt{a_0}\,k_0\,\eta)$. 
Upon 
using the modes~(\ref{eq:hkI}) in the integral~(\ref{eq:cG}), we
obtain the quantity $\cG_{\g\g\g}(\vka,\vkb,\vkc)$ in the first domain
to be
\begin{equation}
\cG_{\g\g\g}^\text{I}({\bf k_1, k_2, k_3}) 
= \frac{k_0}{4\, a_0\,(1+\beta^2)^{3/2}\,\Mpl^3}\,
\f{{\rm e}^{-i\, \alpha\, k_{_{\rm T}}/k_0}}{(k_1\,k_2\,k_3)^{3/2}}\,
\l[\frac{k_0}{\alpha} - \f{\alpha\, k_1\, k_2\, k_3}{k_0\, k_{_{\rm T}}}
+ \frac{i\,k_1\, k_2\, k_3}{k_{_{\rm T}}}\,
\l(\f{1}{k_1} + \f{1}{k_2} + \f{1}{k_3}+ \f{1}{k_{_{\rm T}}}\r)\r],
\label{eq:cGI}
\end{equation}
where $k_{_{\rm T}}= k_1 + k_2 + k_3$.
Similarly, in the second domain, we can make use of the 
solution~(\ref{eq:hkII}) for~$h_k$.
Also, over this domain, we have $z(\eta)=C/\sqrt{a(\eta)}$.
Upon using these expressions and carrying out the integral~(\ref{eq:cG}) 
over $-\alpha\,\eta_0<\eta<\beta\,\eta_0$, we find that 
$\cG_{\g\g\g}(\vka,\vkb,\vkc)$ in the second domain can be written as
\begin{eqnarray}
\cG_{\g\g\g}^{\rm II}({\bf k_1, k_2, k_3}) 
&=& \frac{-i}{4\,\,k_0}\,a_0^2\,(1+\beta^2)^3\,
\Biggl\{A_{\ska}^*\,A_{\skb}^*\,A_{\skc}^*\,
\l({\rm tan}^{-1}\beta + {\rm tan}^{-1}\alpha\r)\nn\\ 
& &+\, \f{1}{6}\, \l(A_{\ska}^*\,A_{\skb}^*\,B_{\skc}^*
+A_{\ska}^*\,B_{\skb}^*\,A_{\skc}^*+B_{\ska}^*\,A_{\skb}^*\,A_{\skc}^*\r)\,
\l[\l(\beta^2 - \alpha^2\r) 
+ 2\;{\rm ln}\,\l(\frac{1 + \beta^2}{1 + \alpha^2}\r)\r]\nn\\
& &+\, \f{1}{135}\,\l(A_{\ska}^*\,B_{\skb}^*\,B_{\skc}^*
+B_{\ska}^*\,A_{\skb}^*\,B_{\skc}^*+B_{\ska}^*\,B_{\skb}^*\,A_{\skc}^*\r)\nn\\
& &\times\,\biggl[3\,\l(\alpha^5 + \beta^5\r)+25\,\l(\beta^3 + \alpha^3\r)
+ 60\,\l(\beta+\alpha\r)- 60\,\l({\rm tan}^{-1}\beta+{\rm tan}^{-1}\alpha\r)\biggr]\nn\\
& &+\,\f{1}{648}\,B_{\ska}^*\,B_{\skb}^*\,B_{\skc}^*\nn\\
& &\times\,\biggl[3\,\l(\beta^8 -\alpha^8\r)+ 32\,\l(\beta^6-\alpha^6\r)
+ 114\,\l(\beta^4-\alpha^4\r) + 96\,\l(\beta^2 - \alpha^2\right) 
- 96\,{\rm ln}\,\l(\f{1+ \beta^2}{1 +\alpha^2}\r)\biggr]\Biggr\}.\label{eq:cGII}
\end{eqnarray}
\end{widetext}
With the aid of the quantities $\cG_{\g\g\g}^{\rm I}({\bf k_1, k_2, k_3})$ 
and $\cG_{\g\g\g}^{\rm II}({\bf k_1, k_2, k_3})$ we have evaluated above 
and, using the solution~(\ref{eq:hkII}) to determine~$h_k(\ef)$, we can 
arrive at the tensor bispectrum by substituting them in the 
expression~(\ref{eq:tbs}).
As the resulting expression proves to be rather lengthy, we do not 
explicitly write down the complete bispectrum here.
In the following section, we shall plot the corresponding tensor 
non-Gaussianity parameter $\hnl$ in the equilateral and the squeezed
limits.


\section{The tensor non-Gaussianity parameter}\label{sec:hnl}

\begin{figure}[!b]
\begin{center}
\includegraphics[width=8.60cm]{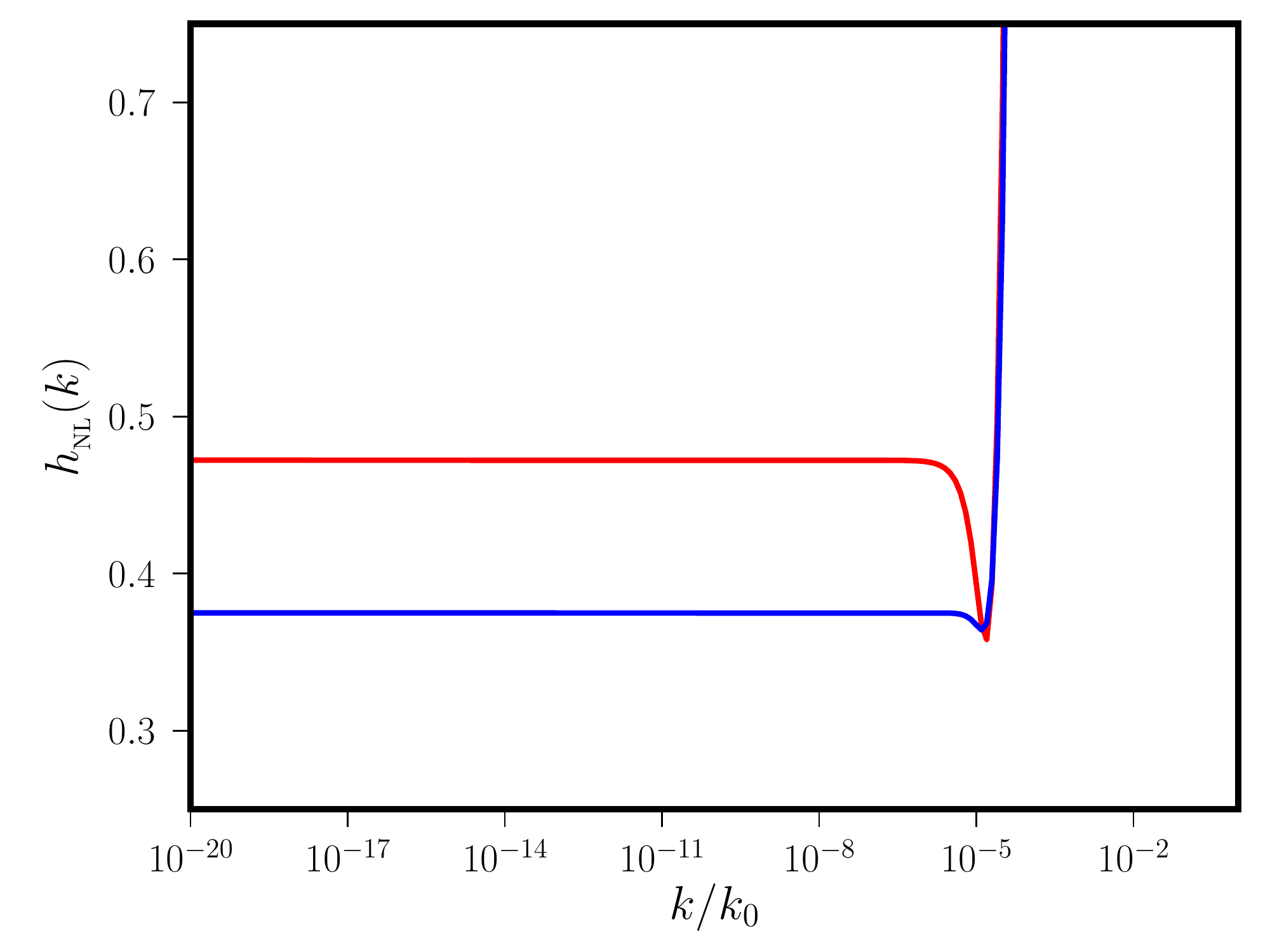} 
\end{center}
\vskip -15pt
\caption{The tensor non-Gaussianity parameter $h_{_{\rm NL}}$ has 
been plotted in the equilateral (in red) and in the squeezed (in 
blue) limits.
We have worked with the same values for the parameters as in the 
previous two figures wherein we had plotted the evolution of the
tensor modes and the corresponding power spectrum.
As in the case of the tensor power spectrum, the analytical results 
are valid for $k/k_0<1/\alpha$, where~$h_{_{\rm NL}}$ is strictly
scale invariant.
In the scale invariant regime, we find that $\hnl=0.472$ and $\hnl=0.375$
in the equilateral and squeezed limits, respectively, which exactly 
correspond to the values under these limits in de Sitter inflation.
In particular, we find that the consistency condition in the squeezed
limit is indeed satisfied.}\label{fig:hnl}
\end{figure}

The dimensionless non-Gaussianity parameter that characterizes the 
amplitude of the tensor bi-spectrum can be defined to be (in this 
context, see, for instance, Refs.~\cite{Sreenath:2013xra,Chowdhury:2015cma})
\begin{widetext}
\begin{eqnarray}
\hnl(\vka,\vkb,\vkc)
&=&-\l(\f{4}{2\,\pi^2}\r)^2\,
\l[k_1^3\, k_2^3\, k_3^3\;G_{\g\g\g}^{m_1n_1m_2n_2m_3n_3}(\vka,\vkb,\vkc)\r]\nn\\
& &\times\,\biggl[\Pi_{m_1n_1,m_3n_3}^{\vka}\,\Pi_{m_2n_2,{\bar m}{\bar n}}^{\vkb}\,
k_3^3\; {\mathcal P}_{_{\rm T}}(k_1)\;{\mathcal P}_{_{\rm T}}(k_2)
+{\rm five~permutations}\biggr]^{-1},\label{eq:hnl-ov}
\end{eqnarray}
where the overbars on the indices imply that they need to be summed 
over all allowed values.
If we ignore the factors involving the polarization tensor, the 
non-Gaussianity parameter $\hnl$ simplifies to
\begin{eqnarray}
\hnl(\vka,\vkb,\vkc)
&=&-\l(\f{4}{2\,\pi^2}\r)^2\, \l[k_1^3\, k_2^3\, k_3^3\; 
G_{\g\g\g}(\vka,\vkb,\vkc)\r]\,
\l[2\,k_3^3\; {\mathcal P}_{_{\rm T}}(k_1)\,
{\mathcal P}_{_{\rm T}}(k_2)
+{\rm two~permutations}\r]^{-1}.\label{eq:hnl}
\end{eqnarray}
\end{widetext}
The non-Gaussianity parameter $\hnl$ in the model of our interest can 
be arrived at using the power and bispectra arrived at in the previous
two sections.
In Fig.~\ref{fig:hnl}, we have plotted the parameter~$\hnl$
over a wide range of wavenumbers in the equilateral and the
squeezed limits. 
If the consistency condition is satisfied, in the squeezed limit
(say, when $k_1\to 0$ and $k_2=k_3=k$) one expects that $\hnl=
(3-n_{_{\rm T}})/8$, where $n_{_{\rm T}}$ is the tensor spectral
index (see, for instance, Ref.~\cite{Sreenath:2014nka}).
For $n_{_{\rm T}}=0$, $\hnl=3/8=0.375$ when the consistency 
condition is satisfied in the squeezed limit, which is exactly 
what we obtain. 
Moreover, in the equilateral limit (say, when $k_1=k_2=k_3=k$), 
we find that $\hnl=0.472$, just as in de Sitter inflation (see, for example, 
Ref.~\cite{Sreenath:2013xra}).
In fact, we find that that the contribution to $\hnl$ due to the
second domain is completely negligible, while the contribution due
to the first domain closely resembles the $\hnl$ that arises in
de Sitter for modes such that $k\ll k_0/\alpha$.
These should be contrasted with the results in a matter bounce in
Einsteinian gravity, wherein $\hnl$ behaves as $k^2$ in the 
equilateral as well as the squeezed limits.
This behavior leads to rather small values for $\hnl$ over scales 
of cosmological interest.
Also, it leads to a violation of the consistency 
condition (in this context, see Ref.~\cite{Chowdhury:2015cma}).


\section{Discussion}\label{sec:d}

The duality principle suggests that bouncing scenarios such as the 
matter bounce can lead to strictly scale invariant spectra as de 
Sitter inflation does~\cite{Wands:1998yp}.
It is expected that the three-point functions can help us discriminate
between alternative scenarios such as the matter bounce and de Sitter 
inflation, which otherwise lead to identical two-point functions.
In inflation, the three-point functions satisfy the consistency condition 
in the squeezed limit, according to which the three-point functions can 
be completely expressed in terms of the two-points functions.
This condition arises due to the fact that the amplitude of the long 
wavelength modes freeze on super-Hubble scales during inflation.
However, in the bouncing models, often, the amplitudes of the modes 
grow strongly as one approaches the bounce, a behavior that results in
the violation of the consistency condition in such scenarios.
It then becomes interesting to examine whether it is possible at all to
restore the consistency condition in the bouncing scenarios.

\par

In this work, we have considered a fairly generic non-minimally coupled 
model of gravitation and examined the behavior of the tensor bispectrum 
in the matter bounce scenario.
We had focused on the case of tensors since it requires limited modeling
and also the modes are easier to determine analytically.
We had explicitly evaluated the tensor bispectrum for a specific form of
the non-minimal coupling function~$G$ [cf.~Eq.~(\ref{eq:nmc})].
We had found that, for such a coupling function, the tensor modes behave 
in exactly the same manner as in de Sitter inflation on sub-Hubble as well 
as super-Hubble scales during the contracting phase.
While there is some difference in the behavior of the tensor modes (when 
compared to their behavior at late times in de Sitter inflation) as they 
approach and cross the bounce, the difference is not substantial enough 
to alter the shape of the tensor power and bispectra.

\par

In retrospect, our choice of the coupling function~(\ref{eq:nmc}) is 
not difficult to understand.
In the Jordan frame wherein gravity is coupled non-minimally to the 
matter fields, it is the function $z(\eta)$ that determines the behavior
of the tensor modes and the resulting correlation functions.
But, in the Einstein frame wherein gravity is coupled minimally, 
the~$z(\eta)$ we have worked with has to be treated as the scale 
factor.
Note that, in the matter bounce scenario of our interest, at early
times during the contracting phase, we have $z(\eta)=C/\sqrt{a(\eta)}
\propto 1/\eta$ [cf. Eq.~(\ref{eq:z})], which is exactly the behavior
of the scale factor in de Sitter inflation.
As we had discussed, it is due to this reason that the modes in 
the first domain $h_k^{\rm I}$ have exactly the same form as in 
de Sitter [cf. Eq.~(\ref{eq:hkI})].
However, it should be emphasized that the equivalence to de Sitter 
inflation is not exact and the equivalence breaks down as one 
approaches the bounce.
In fact, while the amplitude of the tensor modes freeze after leaving 
the Hubble radius during the contracting phase, one finds that there is
a weak growth in their amplitude soon after the bounce [this should be 
clear from the solution~(\ref{eq:hkII})].
Despite this behavior, we find that the domain around the bounce
does not contribute to the tensor bispectrum significantly.
Therefore, the tensor power and bispectra largely retain their forms 
as in de Sitter inflation.
Specifically, we find that our choice of the non-minimal coupling indeed 
restores the consistency condition governing the tensor bispectrum in 
the squeezed limit.

\par

While we have been able to restore the consistency condition, it is
possible that we have achieved it at some cost.
Note that the non-minimal coupling function behaves as $G\propto 
1/\eta^{6}$ [cf. Eq.~(\ref{eq:nmc})] at early stages of the
contracting phase.
This implies that the gravitational coupling to matter is rather strong 
during early times, which can turn out to be undesirable.
We are currently working towards circumventing such difficulties and
examining the corresponding results for the correlation functions involving 
the scalar perturbations in stable contracting phases~\cite{Nandi:2018ooh,
Nandi:2019xlj}.


\section*{Acknowledgements}

LS wishes to thank the Indian Institute of Technology Madras, 
Chennai, India, for support through the Exploratory Research 
Project PHY/17-18/874/RFER/LSRI.


%


\end{document}